\documentclass[aps,prd,floatfix,twocolumn,nofootinbib]{revtex4}
\usepackage{graphicx}
\usepackage{epic}
\usepackage{eepic}
\usepackage{latexsym}

\newcommand{\eq}[1]{(\ref{#1})}
\newcommand{\be}{\begin{equation}}
\newcommand{\ee}{\end{equation}}
\newcommand{\bea}{\begin{eqnarray}}
\newcommand{\eea}{\end{eqnarray}}

\newcommand{\vs}[1]{\vspace{#1 mm}}
\newcommand{\hs}[1]{\hspace{#1 mm}}

\newcommand{\bk}{{\bf k}}
\newcommand{\bx}{{\bf x}}
\newcommand{\bkp}{{\bf k'}}
\newcommand{\bxp}{{\bf x'}}
\newcommand{\bJ}{{\bf J}}
\newcommand{\bK}{{\bf K}}
\newcommand{\bn}{{\bf n}}
\newcommand{\bmm}{{\bf m}}

\def\a{\alpha}
\def\b{\beta}

\def\d{\delta}
\def\D{\Delta}
\def\e{\epsilon}

\def\fr{\frac}

\def\m{\mu}

\def\r{\rho}
\def\s{\sigma}

\def\o{\omega}
\def\dg{\dagger}

\def\del{\partial}

\let\bm=\bibitem
\def\nn{\nonumber}

\begin{document}

\title{Quantum Mechanical Breakdown of Perfect Homogeneity in Reheating After Inflation}

\author{Ali Kaya}
\email[]{ali.kaya@boun.edu.tr}
\affiliation{Bo\~{g}azi\c{c}i University, Department of Physics, \\ 34342,
Bebek, \.Istanbul, Turkey\vs{3}\\Feza G\"{u}rsey Institute,\\
Emek Mah. No:68, \c{C}engelk\"{o}y, \.Istanbul, Turkey\vs{3}}

\date{\today}

\begin{abstract}

In the context of quantum fields in time dependent classical backgrounds, we notice that  the number of created particles with a given momentum largely deviates about its mean value. Guided with this observation we use a complete orthonormal family of localized wave packets to calculate the deviations in the number and energy densities of particles produced in a volume of a given size during reheating. It turns out that at the end of reheating there exists (in general tiny) spatial variations in these densities on Hubble length scales over which local interactions are incapable of restoring homogeneity. This signals the destruction of perfect homogeneity attained after inflation due to the quantum nature of particle production process in reheating. 

\end{abstract}

\maketitle

\section{Introduction}

In scalar field driven inflationary models, the universe expands (almost) exponentially due to the scalar potential energy density acting as an effective cosmological constant. As a result, a small pre-inflationary causal patch enlarges to encompass the whole observed universe. With this notion,  inflation solves many puzzles of the standard cosmological model like homogeneity, isotropy and monopole problems.

As inflation ends, one finds an extremely smooth, flat universe which is practically at {\it zero} temperature and filled only with a coherently oscillating inflaton field of decreasing amplitude about the minimum of its potential. Almost all matter populating the universe in the subsequent radiation era should have been produced by the decay of the inflaton field. These decay products later become thermalized by collisions and further decays and this marks the beginning of the usual hot stage.

To have a complete cosmic history of the universe, it is crucial to understand reheating after inflation in detail. The elementary theory of reheating (for a review see e.g. \cite{erh1}) is based on perturbation theory which has obvious limitations. More recently, the importance of the parametric resonance effects on the decay of the inflaton field is recognized in \cite{pr1,pr2,pr3,y1,y2,y3,kls1, kofrev, kls2} (for a recent review see, e.g. \cite{rev}). Especially it is shown in \cite{kls1} that in many models reheating actually starts by a decay due to broad parametric resonance, called preheating. The particle creation effects in the broad parametric resonance regime is studied in \cite{kls2}, which shows that  the whole process becomes stochastic due to the expansion of the universe but one  still gets exponentially growing  occupation numbers. 

Consider the universe just after the inflation at the beginning of the reheating stage, which has already acquired a huge size. Viewing this moment as the {\it starting point} of the subsequent evolution, causality requires that the particle creation and thermalization processes occur independently in each succeeding Hubble volume, which (ignoring the slight difference between the horizon and the Hubble distances) form causally disjoint regions during reheating. Naively thinking, this does not imply a breakdown in homogeneity since in each region we have identical copies of the same oscillating scalar field. However, since the decay process is quantum mechanical in nature, the number of created particles in each volume is expected to fluctuate about a mean value which may induce (in general small)  density inhomogeneities. The aim of this work is to calculate these variations by analyzing the decay process of the inflaton field using localized wave packets. 

The plan of the paper is as follows. In the following section, we review the well known properties of the quantum mechanical harmonic oscillator with a time dependent frequency. We point out some salient features of this system which will be important in the following sections. In section \ref{sec3}, we study the general framework of particle production by time dependent external classical fields using a complete orthonormal family of localized wave packets and determine the deviations in the number and the  energy densities of particles produced in a given volume. In section \ref{sec4} , we apply these results to a generic inflationary scenario and estimate the amount of inhomogeneities. We conclude with a brief review of our findings in \ref{sec5}. 

\section{Harmonic Oscillator with Time Dependent Frequency}\label{sec2}

In this section we review the basic properties of the quantum mechanical harmonic oscillator with time dependent frequency $\o$. As it is well known, this system mimics the essential features of the particle creation process during reheating. For a unit mass particle, the Hamiltonian of the oscillator is given by 
\be
H=\fr{1}{2}\left[p^2+\o^2q^2\right],
\ee
where $q$ and $p$ denote the position and the momentum operators obeying 
\be\label{cc} 
[q,p]=i.
\ee
In the Heisenberg picture these operators obey 
\bea
\dot{q}=p,\hs{5}\dot{p}=-\o q,
\eea
where dot denotes time derivative. To solve the system one may introduce time dependent ladder operators $a$ and $a^\dagger$ via 
\be
q=\fr{a}{\sqrt{2\o}}e^{-i\int^t \o dt}+\fr{a^\dagger}{\sqrt{2\o}}e^{i\int^t\o dt}.
\ee
In taking time derivative of $q$, one wishes to treat $a$, $a^\dagger$ and $\o$ as if they were time independent quantities (like in the usual harmonic oscillator) so that 
\be
p=\dot{q}=\fr{-i\o a}{\sqrt{2\o}}e^{-i\int^t \o dt}+\fr{i\o a^\dagger}{\sqrt{2\o}}e^{i\int^t\o dt}.
\ee
This can be achieved by imposing 
\be\label{ae}
\dot{a}=\fr{\dot{\o}}{2\o}e^{2i\int^t \o dt}a^\dg,\hs{5}
\dot{a}^\dg=\fr{\dot{\o}}{2\o}e^{-2i\int^t \o dt}a.
\ee
The nice feature of defining the ladder operators in this fashion is that the canonical commutation relation \eq{cc} gives  
\be\label{cca}
[a,a^\dg]=1,
\ee
and the Hamiltonian becomes
\be
H=\o\,\left[a^\dagger a +\fr{1}{2}\right].
\ee
One can then construct the Hilbert space as usual; the instantaneous ground state of the system at time $t$ is defined by 
\be\label{d1}
a|0_t>=0,
\ee
and an orthonormal basis of energy eigenvectors can be found as 
\be
|n>=\fr{(a^\dg)^n}{\sqrt{n!}}|0_t>.
\ee

The operator equation \eq{ae} can be solved in terms of the constant operators at time $t_0$, i.e. $a_0$ and $a_0^\dagger$, by introducing a Bogoligov  transformation 
\bea
a&=&\a\,a_0 + \b^*\, a_0^\dg,\nn\\
a^\dg&=& \b\,a_0+ \a^*\, a_0^\dg.\label{bog}
\eea
The complex functions $\a$ and $\b$ obey the same differential equations as $a$ and $a^\dagger$ given in \eq{ae}, respectively. Eq. \eq{cca} requires
\be
|\a|^2-|\b|^2=1,
\ee
and initially one should choose $\a(t_0)=1$, $\b(t_0)=0$. 

The ground state of the system at time $t_0$, which is defined by  
\be\label{d2}
a_0|0_0>=0,
\ee
will not remain to be the ground state at a later time. Rather, the expectation value of the number operator $N=a^\dagger a$ in this state can be found as
\be\label{qn}
<N>=<0_0|N|0_0>=|\b|^2,
\ee
which shows that {\it on the average} this state contains $|\b|^2$ quanta at time $t$. One can actually be more precise  and determine $|0_0>$ exactly. Expanding it in the orthonormal basis vectors \label{orb} and using \eq{bog} and \eq{d2} it can be shown  that up to an irrelevant phase 
\be\label{00}
|0_0>=\fr{1}{\sqrt{|\a|}}\exp(\fr{\b^*}{2\a^*}a^\dg a^\dg) |0_t>.
\ee
From this expression one can read of the probability $P_{2n}$ of finding an even number of quanta in $|0_0>$
\be\label{pd}
P_{2n}=\fr{(2n)!}{2^{2n}(n!)^2}\fr{|\b|^{2n}}{|\a|^{2n+1}},
\ee
where the numerical prefactor can be recognized as the $n$'th order Taylor expansion coefficient of $1/\sqrt{1-x}$. It is easy to see that $P_{2n}$ is a decreasing function of $n$ and for $n\gg1$ one has 
\be
P_{2n}\simeq\fr{1}{\sqrt{\pi n}}\fr{|\b|^{2n}}{|\a|^{2n+1}}.
\ee
Although on the average $|0_0>$ contains $|\b|^2$ quanta, {\it the most probable} outcome of a measurement is the ground state $|0_t>$ with no quanta. 

To characterize the probability distribution one can calculate the deviation $\D N$ of the number of quanta 
\be
(\D N)^2\equiv<N^2>-<N>^2.
\ee
Using 
\be
<N^2>=<0_0|N^2|0_0>=2|\a|^2|\b|^2+|\b|^4
\ee
one finds 
\be\label{dn}
\fr{\D N}{<N>}=\sqrt{2}\,\fr{|\a|}{|\b|}>\sqrt{2}.
\ee
For large average production, i.e. $|\b|\gg1$, the relative deviation is equal to $\sqrt{2}$. In the opposite limit of small quanta creation, i.e. $|\b|\ll1$, it is given by $\sqrt{2}/|\b|$, which is much larger than unity. In any case, the main conclusion here is that the number of created quanta has large fluctuations about the mean value given by \eq{qn}. 

\section{Quantum Particle Production: A Wave Packet Analysis}\label{sec3}

Consider a real  scalar field $\chi$ propagating in a cosmological Robertson-Walker background 
\be
ds^2=-dt^2+a^2(dx^2+dy^2+dz^2),
\ee
which has the action
\be
S=-\fr12 \int \sqrt{-g}\left[(\nabla\chi)^2+M^2 \chi^2\right].
\ee
We assume that in addition to the scale factor $a$, the mass parameter $M$ may also depend on time $t$. Defining a new field by
\be
X=a^{3/2} \chi,
\ee
the action up to surface terms becomes 
\bea\nn
S=\fr12 \int \left[\dot{X}^2-\fr{(\del_i X)^2}{a^2}-(M^2-\fr94 H^2 -\fr32 \dot{H}) X^2\right],
\eea
where dot denotes time derivative and $H=\dot{a}/a$ is the Hubble parameter. One can expand the field in Fourier modes and introduce ladder operators as in the harmonic oscillator system discussed in the previous section 
\bea
X=\fr{1}{(2\pi)^{3/2}} \int d^3 k\left[\fr{a_\bk}{\sqrt{2\o_k}}e^{-i\bk.\bx-i\int^t\o_k dt}+h.c.\right],\nn
\eea
where h.c. denotes hermitian conjugate, $\bk$ is the comoving wave vector, $k^2=k^ik^j\d_{ij}\equiv\bk.\bk$  and 
\be\label{o}
\o_k^2=M^2+\fr{k^2}{a^2}-\fr94 H^2 -\fr32 \dot{H}.
\ee
The time dependence of the operators $a_\bk$ and $a_\bk^\dg$ are imposed to get the conjugate momentum $\Pi=\dot{X}$ as
\bea\nn
\Pi=\fr{1}{(2\pi)^{3/2}} \int d^3 k\left[\fr{-i\o_k a_\bk}{\sqrt{2\o_k}}e^{-i\bk.\bx-i\int^t \o_k dt}+h.c.\right],
\eea
which implies
\bea
\dot{a}_\bk&=&\fr{\dot{\o_k}}{2\o_k}e^{2i\int^t \o_k dt}a_{-\bk}^\dg,\nn\\
\dot{a}_{-\bk}^\dg&=&\fr{\dot{\o_k}}{2\o_k}e^{-2i\int^t \o_k dt}a_\bk.\label{aak}
\eea
In this case the equal time canonical commutation relation $[X,\Pi]=i\d(\bx-\bxp)$ is equivalent to 
\be\label{fcc}
[a_\bk,a^\dg_\bkp]=\d(\bk-\bkp)
\ee
and the Hamiltonian can be expressed as
\be\label{fh}
H=\int d^3 k\left[a_\bk^\dg a_\bk +\fr{1}{2}\right]\,\o_k.
\ee
The instantaneous ground state is defined by 
\be\label{gs}
a_\bk |0_t>=0,
\ee
and the Hilbert space can be build by acting with the creation operators $a_\bk^\dg$ on $|0_t>$.

One thus rediscovers the well-known fact that the free field theory of $\chi$ particles is nothing but an infinite collection of harmonic oscillators each of which is labeled by a comoving wave vector $\bk$ and by the time dependent frequency \eq{o}. The only (minor) complication is that due to conservation of momentum ladder operators having wave vectors $+\bk$ and $-\bk$ are coupled by \eq{aak}. 

To solve \eq{aak} one can introduce a Bogoligov transformation 
\bea
a_\bk&=&\a_k a_\bk(t_0) +\b_k^* a_{-\bk}^\dg(t_0),\nn\\
a_{-\bk}^\dg&=&\b_k a_\bk(t_0) +\a_k^* a_{-\bk}^\dg(t_0),.
\eea
where 
\bea
\dot{\a}_k&=&\fr{\dot{\o_k}}{2\o_k}e^{2i\int^t \o_k dt}\b_k,\nn\\
\dot{\b}_k&=&\fr{\dot{\o_k}}{2\o_k}e^{-2i\int^t \o_k dt}\a_k. \label{be}
\eea
Note that $\a_k$ and $\b_k$ depend only on the magnitude $k$ and not on the direction. 

As in the harmonic oscillator problem, the ground state of the system at time $t_0$,
\be\label{gs0}
 a_{\bk}(t_0)|0_0>=0,
\ee
becomes a multiparticle state at a later time. Similar to \eq{00}, it can be expressed in terms of the ground state $|0_t>$ at time $t$ as
\be
|0_0>=\prod_{\bk}\fr{1}{\sqrt{|\a_k|}}\exp(\fr{\b_k^*}{2\a_k^*}a_{-\bk}^\dg a_\bk^\dg)\, |0_t>,
\ee
i.e. for each momentum mode labeled by $\bk$ the probability distribution for the number of such particles contained in $|0_0>$ is given by \eq{pd}, where half of the particles have momentum $\bk$ and the other half have $-\bk$. Note however that as it stands the above formula does not make sense and needs regularization. 

The mean number of modes contained in $|0_0>$ can be found by calculating the expectation value of the number operator 
\be\label{an}
N_\bk=a^\dg_\bk a_\bk,
\ee
which  by \eq{fcc} reads 
\be\label{fn}
<N_\bk>=<0_0|N_\bk|0_0>=|\b_k|^2 \d({\bf 0}).
\ee
The infinity contained in the delta function can be interpreted as
\be
\d({\bf 0})=\fr{V}{(2\pi)^3},
\ee
where $V$ is the total comoving volume of the space. Defining the comoving number density 
\be
n_\bk=\fr{N_\bk}{V},
\ee
one obtains the familiar expression 
\be\label{nn}
<n_\bk>=<0_0|n_\bk|0_0>=\fr{|\b_k|^2}{(2\pi)^3},
\ee
which is well defined. 

Although it looks like one manages to make sense of the infinity in \eq{fn}, there is still an interpretation problem. The momentum modes  we are dealing with are completely dislocalized and it is not meaningful to talk about the density of such excitations. Moreover, the calculated mean values should be explained in an ensemble picture but in our case there is only one realization of the system, i.e. in a given spacetime one can only make a single measurement on the mode labeled by $\bk$. As pointed out in the previous section, the result of a measurement may not be close to the average value  since the deviation from the mean is large. Indeed the most probable outcome of a single measurement is the state with no particle. 

There is an alternative and physically more viable way of dealing with the above problem. The infinity appeared in \eq{fn} can be thought to arise due to the particle creation in an infinitely large space. A similar divergence also appears in the particle creation by black holes which was interpreted as the finite steady rate of emission for an infinite time as shown in  \cite{haw} by analyzing the process using 
localized wave packets. Following \cite{haw} we introduce a new set of ladder operators as follows. Let $\e>0$ be an arbitrary comoving momentum scale and introduce two vectors $\bJ=(j_1,j_2,j_3)$ and $\bn=(n_1,n_2,n_3)$ with {\it integer} entries. Let 
\be\label{ajn}
a_{\bJ\bn}= \fr{1}{\e^{3/2}}\int_{\bJ}\,\exp(-2\pi i\bn .\bk/\e)\,a_\bk,
\ee
where we introduce a shorthand notation for a three dimensional momentum integral  
\be
\int_{\bJ}\equiv \int_{j_1\e}^{(j_1+1)\e} \int_{j_2\e}^{(j_2+1)\e}\int_{j_3\e}^{(j_3+1)\e}\,dk_1 dk_2 dk_3. 
\ee
As we show below, $a_{\bJ\bn}$ is the annihilation operator for a mode peaked around the comoving position $2\pi\bn/\e$ with a spread $1/\e$ in each direction. For small $\e$, the mode can be thought to have a comoving momentum $\bJ\e$. The new operators obey 
\be\label{acr}
[a_{\bJ\bn},a^\dg_{\bK\bmm}]=\d_{\bJ\bK}\d_{\bn\bmm}, 
\ee
which justifies the identification of $a_{\bJ\bn}$ and $a^\dg_{\bJ\bn}$ as ladder operators 

It is possible to invert \eq{ajn} and express $a_\bk$  in terms of $a_{\bJ\bn}$ as
\be
a_\bk=\fr{1}{\e^{3/2}}\sum_\bn \exp(2\pi i\bn .\bk/\e)\, a_{\bJ\bn};\hs{3} j_i\e<k_i<(j_i+1)\e.
\ee
Using this expression in the Fourier expansion of $X$ one gets
\be\label{xx}
X=\sum_{\bJ,\bn}\left[f_{\bJ\bn}({\bf x})\fr{a_{\bJ\bn}}{\sqrt{2\o_k}}e^{-i\int^t\o_k dt}+h.c.\right],
\ee
where the mode functions are given by
\be
f_{\bJ\bn}({\bf x})=\fr{1}{(2\pi\e)^{3/2}}\int_\bJ \exp(-i\bk.({\bf x}-2\pi\bn/\e)).
\ee
Performing the integral it is easy to see that $f_{\bJ\bn}$ is localized around ${\bf x}=2\pi\bn/\e$ with width $1/\e$. Moreover
\be
\int d^3 x\, f_{\bJ\bn}\, f^*_{\bK\bmm} = \d_{\bJ\bK}\,\d_{\bn\bmm},
\ee
and 
\be
\sum_{\bJ,\bn} f_{\bJ\bn}({\bf x})\, f^*_{\bJ\bn}({\bf x'})=\d({\bf x}-{\bf x'}),
\ee
which imply  that these functions form a complete orthonormal family. From the expansion \eq{xx} we see that the operator $a^\dg_{\bJ\bn}$ acting on the ground state produces a localized field quanta with the wavefunction  $f^*_{\bJ\bn}$.

Ignoring the zero-point energy, the Hamiltonian \eq{fh} can also be expressed in terms of the new ladder operators as
\be\label{ah}
H=\fr{1}{\e^3}\sum_{\bJ,\bn,\bmm}\int_\bJ\o_k\exp(2\pi i\bk.(\bn-\bmm)/\e)\,a^\dg_{\bJ\bn}\,a_{\bJ\bmm}.
\ee
Therefore the Fock space states
\be\label{afs}
a^\dg_{\bJ\bn}...a^\dg_{\bK\bmm}|0_t>
\ee
are {\it not}  the eigenvectors of the Hamiltonian. Note that the definition of the ground state in \eq{gs} is equivalent to $a_{\bJ\bn}|0_t>=0$.  

If $\o_k$ does not change significantly in the interval $j_i\e<k_i<(j_i+1)\e$ for a given $\bJ$, then it may be taken out of the integral in \eq{ah}. In this case, the momentum integral yields $\e^3\d_{\bn\bmm}$ and the Hamiltonian becomes diagonalized in this particular subspace of basis vectors.  If this condition is satisfied for all $\bJ$ then all the Fock space states become (approximate) eigenvectors of the Hamiltonian. 

Mathematically speaking, $\o_k$ can be treated as a constant in \eq{ah} if  $\e$ is small enough such that 
\be\label{ck}
\o_k\gg\e\, d\o_k/dk.
\ee
For a massive field this condition is satisfied for all momenta if $\e\ll m$ and for a massless field it is only obeyed in the momentum range $k\gg\e$. Assuming \eq{ck} holds, the Hamiltonian turns into 
\be
H\simeq\sum_{\bn}\left[\sum_{\bJ} \o_{j\e} \,a^\dg_{\bJ\bn}\,a_{\bJ\bn}\right],
\ee
where $j^2=\bJ.\bJ$. Recalling that the sum over $\bn$ can be viewed as a sum over cubic regions of side length $2\pi/\e$, one can define the total number and the energy density operators as
\bea
n_\e&\equiv&\fr{\e^3}{(2\pi)^3} \sum_\bJ \,a^\dg_{\bJ\bn}\,a_{\bJ\bn},\label{ne}\\
\r_\e&\equiv&\fr{\e^3}{(2\pi)^3} \sum_\bJ \o_{j\e}\,a^\dg_{\bJ\bn}\,a_{\bJ\bn}.\label{re}
\eea
In analogy with \eq{an}, one can also introduce the number operator for a specific mode 
\be
N_{\bJ\bn}=a^\dg_{\bJ\bn}\,a_{\bJ\bn}.
\ee
Note that by \eq{acr}  $N_{\bJ\bn}$ is an honest counting operator. 

Let us now analyze particle creation effects using these localized modes. The expectation value of   $N_{\bJ\bn}$ in the ground state \eq{gs0} can be calculated as
\be\label{fn2}
< N_{\bJ\bn}>=<0_0| N_{\bJ\bn} |0_0>=\fr{1}{\e^3}\int_\bJ |\b_k|^2. 
\ee
Dividing with the volume of the region one gets the number density as
\be\label{nn2}
< n_{\bJ\bn}>=\fr{1}{(2\pi)^3}\int_\bJ |\b_k|^2. 
\ee
Although \eq{fn2} and \eq{nn2}  look like \eq{fn} and \eq{nn}, respectively, it is a lot easier to interpret them physically. First of all, since we are dealing with localized excitations  it is meaningful  to talk about  the number density of modes. Secondly, unlike the previous expression \eq{fn}, \eq{fn2}  is free of  infinities. Finally, even in a single realization, i.e. in a given spacetime, it is possible to make sense of the mean values since one can make independent measurements in different spatial regions  labeled by $\bn$  and calculate the  average of the outcomes. Note that due to translational invariance the mean values \eq{fn2} and \eq{nn2} do not depend on $\bn$. 

As discussed in the previous section, an important property of the particle creation effects in this setup is the existence of large deviations about the average values. This clearly indicates a breakdown in the homogeneity since the number of particles with a fixed quantum number $\bJ$  produced in different regions labeled by $\bn$ will substantially be different. 

There are, however, two points one should be careful about before concluding that the homogeneity is destroyed. Firstly, the interactions between particles work for homogenization. Since the Hubble distance sets a limit for the range of local interactions in an expanding universe, these can be neglected  if $\e$ is chosen to have this scale. Secondly, for small $\e$ the spacing between neighboring  momentum levels is narrow and statistics might work to reduce the deviation by averaging over nearby levels. To take into account this point, one should calculate the deviations of the total number of modes $n_\e$ or of the energy density $\r_\e$, defined in \eq{ne} and \eq{re}.

To determine those deviations we first note that straightforward manipulations yield
\be
<n_\e>=<0_0|n_\e|0_0>=\fr{1}{(2\pi)^3}\int d^3 k |\b_k|^2
\ee
and 
\bea
&&(\D n_\e)^2=<0_0|n_\e^2|0_0>-(<0_0|n_\e|0_0>)^2\nn\\
&&=\fr{1}{(2\pi)^6}\sum_\bJ\left(\int_\bJ |\b_k|^2 \int_\bJ |\a_{k'}|^2 
+\int_\bJ \b_k\a_k^* \int_\bJ \b_{k'}^* \a_{k'} \right).\nn
\eea
For small $\e$ obeying $|\b_k|\gg\e\, d|\b_k|/dk$, the integrals can be approximated, e.g. by
\be
\int_\bJ |\b_k|^2\simeq \e^3 |\b_{j\e}|^2.
\ee
Collecting terms and converting again sums into integrals one obtains the final result for the relative deviation 
\be\label{m1}
\fr{\D n_\e}{<n_\e>}=\sqrt{2}\,\e^{3/2}\,\fr{\left(\int d^3 k |\b_k|^2|\a_k|^2\right)^{1/2}}{\int d^3 k |\b_k|^2}.
\ee
Similarly the deviation in the energy density defined in \eq{re} can be found as 
\be\label{m2}
\fr{\D \r_\e}{<\r_\e>}=\sqrt{2}\,\e^{3/2}\,\fr{\left(\int d^3 k |\b_k|^2|\a_k|^2\o_k^2\right)^{1/2}}{\int d^3 k |\b_k|^2\o_k}.
\ee
Let us recall that normalization requires $|\a_k|^2-|\b_k|^2=1$.  

It is possible to understand the origin of $\e$ dependence in \eq{m1} and \eq{m2} as follows. In a homogeneous background the number of modes in a volume $V$ is proportional to $V$ and statistically the relative deviation of a total quantity is expected to decrease with the square root of the number of modes, i.e. like $1/\sqrt{V}$. In our case $V\sim1/\e^3$ which explains $\e$ dependence in the above  formulas.  The deviation in each excitation labeled by $\bJ$ can be much larger by \eq{dn}, and thus  \eq{m1} and \eq{m2} set a {\it lower bound} in the degree of  inhomogeneities on the comoving scale $\e$ which can only be saturated if local interactions are very efficient on that scale. 

To determine the power spectrum $P(k)$ corresponding to \eq{m2}, a Gaussian or top-hat window function can be introduced to probe the scale $k\sim \e$ (see, e.g., \cite{kt}). A simple calculation shows that \eq{m2} has a white noise spectrum similar to sub-horizon thermal fluctuations studied in \cite{thermal}.

\section{Application to Reheating}\label{sec4}

In this section, we consider the particle creation effects during reheating in a scalar field driven inflationary model. The evolution of the background fields, i.e. the metric and the inflaton, is governed by the Einstein and the scalar field equations 
\bea
&&H^2=\fr{8\pi}{3M_p^2}\left[\fr12 \dot{\phi}^2+V(\phi)\right],\nn\\
&&\ddot{\phi}+3H\dot{\phi}+\fr{\del V}{\del\phi}=0,
\eea
where $V(\phi)$ is the scalar potential. In some classes of inflationary models, e.g. in chaotic inflation, the scalar oscillates about the minimum of the potential during reheating, so one can take 
\be\label{pot}
V=\fr12 m^2\phi^2,
\ee
where $m$ is the inflaton mass. In a chaotic inflationary scenario $V$ can be assumed to have exactly this form. Due to the expansion of the universe the amplitude of the oscillations gradually decreases in time so one can write the solution for the inflaton as 
\be
\phi=\Phi(t)\,\sin(mt).
\ee
Assuming $\dot{\Phi}\ll m\Phi$, the Einstein and the scalar equations become
\be\label{sol}
H^2=\fr{4\pi m^2}{3M_p^2} \Phi^2,\hs{5}\dot{\Phi}+\fr32 H\Phi=0,
\ee
which can be solved as
\be\label{sol1}
a=a_0\left(\fr{t}{t_0}\right)^{2/3},\hs{5}\Phi=\fr{M_p}{\sqrt{3\pi}mt}.
\ee
As it is well known, the stress-energy-momentum tensor corresponding to  the coherent inflaton oscillations is equivalent to the one for the pressureless dust.  In this case, the combination $9H^2/4+3\dot{H}/2$ which appears in \eq{o} is equal  to zero. 

We first consider particle creation effects during preheating  in a chaotic inflationary scenario with a quadratic potential \eq{pot}. This has been studied  in detail both analytically and numerically in \cite{kls2} and we mainly use the findings of that paper below. In such a model inflation occurs when $\phi\geq M_p$ and it ends when the field decreases below $\phi\sim M_p/2$.  It turns out that even after a single oscillation the amplitude drops enormously and the solution \eq{sol1} becomes a very good approximation. Following \cite{kls2}, we take $t_0=\pi/2m$ as the time for the beginning of preheating and thus initially 
\be
\Phi_0=\fr{2M_p}{\pi\sqrt{3\pi}}\simeq\fr{M_p}{5},\hs{5}H_0=\fr{4m}{3\pi}. 
\ee
A realistic value for the inflaton mass is $m=10^{-6}M_p$. 

We assume that inflaton is coupled to a light boson $\chi$ with an interaction term 
\be\label{int1}
{\cal L}_{int}=-\fr12 g^2 \phi^2\chi^2,
\ee
where $g$ is a dimensionless coupling constant usually assumed to be small. In this case, the results of the previous section can be applied  by fixing  the time varying mass in \eq{o} as $M^2=g^2\Phi^2\sin^2(mt)$. As shown in \cite{kls2}, for $g\gg10^{-6}$ one finds a decay due to broad parametric resonance and the occupation numbers of the modes which have comoving momenta less than $k_*$ are exponentially growing, where
\be
\fr{k_*}{a_0}=\sqrt{gm\Phi_0}.
\ee

The particle creation effects in this model are characterized by an effective index $\m_k$ such that 
\be
\b_k=e^{\m_k m t}.
\ee
The index can be estimated  as
\be\label{ind}
\m_k\simeq\m-\fr12 \m_k''(k_m)\,(k-k_m)^2,
\ee
where the second derivative of $\m_k$ can be approximated by $\m_k''(k_m)=2\m/\D k$. Here $k_m$ and $\D k$, which are the maximum and the width of the first and the most important resonance band, are nearly equal to $k_*/2$. Depending on $g$ the average index $\m$ varies between $0.1$ and $0.2$ and for numerical estimations one can take $\m\sim 0.13$ \cite{kls2}. 

Assume that the resonance ends at time $t_1$ after $N$ oscillations, where $mt_1=2\pi N$. From \eq{sol} and \eq{sol1} the Hubble constant at the end of resonance can be found as  $H_1=m/(3\pi N)$.  In calculating the relative deviations, we choose $\e$ to be the comoving Hubble scale at the end of preheating, i.e. 
\be
\e=a_1H_1.
\ee
It is easy to see that in the broad resonance regime $\e\ll k_*$ and the condition \eq{ck} is satisfied. 

Using \eq{ind}, the integrals in \eq{m1} and \eq{m2} can be evaluated by the steepest decent method \cite{kls2}. Nothing that in this regime $\a_k\simeq\b_k$, one can straightforwardly find
\be\label{fd}
\fr{\D \r_\e}{<\r_\e>}\simeq\fr{\D n_\e}{<n_\e>}\simeq\fr{2^{9/4}}{3^{3/2}\pi^2}\,\left(\fr{\m}{N}\right)^{1/4}\,\left(\fr{g\Phi_0}{m}\right)^{-3/4}.
\ee
For small $g$, the resonance may end before back reaction and rescattering effects become important. The number of oscillations for the first stage, where these effects can be ignored, are approximately given by $N\simeq gM_p/(6\pi m)$ \cite{kls2}. Usually, the time for the second stage is much more shorter so one can take $N$ to have this value in estimating the deviation \eq{fd}. Setting $\m=.13$ and $m=10^{-6}M_p$ one finally obtains 
\be\label{fd2}
\fr{\D \r_\e}{<\r_\e>}\simeq\fr{\D n_\e}{<n_\e>}\simeq\fr{2\times 10^{-7}}{g}.
\ee
For $g\ll3\times 10^{-4}$ resonance ends in the first stage and the relative deviation is much larger than $10^{-3}$. For bigger  $g$, one should consider back reaction and rescattering effects and the deviation \eq{fd2} gives the degree of inhomogeneity in the beginning of this stage. Note that even for $g=10^{-2}$ the relative deviation has the same order of magnitude with the fluctuations in the CMB temperature.  

As pointed out above, \eq{fd} actually gives a lower bound since the deviations in the numbers of individual excitations can be much larger.  By choosing $\e$ to be the Hubble scale at the end of preheating we make sure that local interactions cannot restore homogeneity. 

Depending on the details of the theory, the reheating process may continue after preheating but the decay of the inflaton ceases to exist in the parametric resonance channel. In this period, the well known perturbation theory of reheating is applicable to study the particle creation effects. We now continue with the determination of relative deviations in this regime in a different model. 

In the following we assume a trilinear coupling, which may arise after spontaneous symmetry breaking,
\be\label{int2}
{\cal L}_{int}=-\fr12 \s \phi\chi^2,
\ee
where $\s$ is a constant with mass dimension. As shown in \cite{dfkpp}, the  preheating picture completely changes when both interections \eq{int1} and \eq{int2} present in the Lagrangian. Therefore, the perturbative decay due to \eq{int2} should be considered on its own, i.e. it is not to be preceded by the preheating considered above. Our aim here is to determine fluctuations in a well-known perturbative decay scenario based on \eq{int2}. 

The particle creation process via trilinear interaction  was outlined in \cite{kofrev}. The perturbation theory is applicable when $\s\Phi/m^2\ll1$. Note that the evolution of the background fields is still given by \eq{sol1}, however the amplitude $\Phi$ is now small compared to its magnitude in the parametric resonance regime. The frequency \eq{o} corresponding to the interaction \eq{int2} is given by 
\be
\o_k^2=\fr{k^2}{a^2}+\s\Phi\sin(mt).
\ee
Due to the expansion equivalent to matter domination, $H$ dependent terms in \eq{o} cancel each other. 

For $|\b_k|\ll1$, an iterative solution to \eq{be} is given by   
\be\label{ps}
\b_k\simeq \fr12 \int_{t_0}^{t_1}dt\, \fr{\dot{\o}_k}{\o_k}\exp\left(-2i\int^t \o_k(t')dt'\right),
\ee
where $t_0$ and $t_1$ denote the beginning and ending of the decay, respectively. This integral can be evaluated using the stationary phase method \cite{strqc}.  It is easy to see that $\dot{\o}_k/\o_k$ term in \eq{ps} consists of a non-oscillatory ignorable piece together with a term proportional to $\cos(mt)$. Since the phase in \eq{ps} is negative definite, only $e^{+imt}$ part contributes to the integral. As a result one finds
\be
\b_k\simeq\fr18 \int_{t_0}^{t_1}\,\fr{m\s\Phi}{\o_k^2}\, e^{if(t)}\,dt,
\ee 
where the phase function is
\be
f(t)=\int^t\left[-2 \o_k(t')+m\right]dt'.
\ee
For a given $k$, the main contribution to the integral comes from an interval near  $t_*$ fixed by $\o_k(t_*)=m/2$, which, in the perturbative regime $\s\Phi/m^2\ll1$, implies 
\be\label{c1}
\fr{k}{a_*}=\fr{m}{2}.
\ee
This corresponds to the decay of the inflaton at time $t_*$ to two $\chi$ particles with comoving momentum $k$. 

To apply the stationary phase approximation we first note that $\ddot{f}(t_*)=mH(t_*)-2\s\Phi(t_*)\cos(mt_*)$. From \eq{sol} and \eq{sol1}, the oscillating factor can be ignored for $m^2/M_p>\s$ which yields  
\bea
\b_k\simeq\fr{e^{if(t_*)}\s\Phi(t_*)}{2m} \int_{t_0}^{t_1}\, \exp\left[imH(t_*)(t-t_*)^2/2\right]\,dt.\nn
\eea
Since $m\gg H$, the stationary phase approximation is valid and the limits of the integral can be extended  to infinity. This up to an irrelevant phase gives 
\be
\b_k\simeq\fr{\s}{m^2}\sqrt{\fr{\pi\Phi(t_*)M_p}{2}}.
\ee
To determine $\Phi(t_*)$, we note that \eq{sol1} implies
\be
\Phi(t_*)=\Phi_0\,\left(\fr{a_0}{a_*}\right)^{3/2}=\Phi_0\,a_0^{3/2}\,\left(\fr{m}{2k}\right)^{3/2},
\ee
where in the last line we use \eq{c1} to determine  $a_*$ in terms of $k$. This gives  
\be\label{bf}
|\b_k|^2\simeq\fr{\pi\s^2M_p\Phi_0}{2m^{5/2}}\left(\fr{a_0}{2k}\right)^{3/2},
\ee
where the comoving momentum $k$ should be in the decay range 
\be\label{dr}
\fr{a_0m}{2}<k<\fr{a_1m}{2}
\ee
with $a_1$ being the scale factor at time $t_1$. In the perturbative regime $\s\Phi/m^2\ll1$ (and for $m^2/M_p>\s$), $|\b_k|\ll1$ which justifies both the solution \eq{ps} and the stationary phase approximation used in the calculation of the integral. 

The whole process ends when the energy density $\rho_\chi$ of $\chi$ particles cathes up the energy density of the inflaton, i.e.
\be\label{ce}
\rho_\chi\equiv\fr{1}{(2\pi a_1)^3}\,\int d^3 k \,|\b_k|^2=\fr12 m^2\Phi_1^2,
\ee
where $\Phi_1$ is the amplitude of the oscillations at the end of the decay. Using \eq{bf} and recalling that the integral in the momentum space is defined  in the spherical region \eq{dr}, we find 
\be\label{er}
\rho_\chi\sim\s^2M_p\Phi_1\left[1-\left(\fr{a_0}{a_1}\right)^{5/2}\right].
\ee
For $a_1\gg a_0$, \eq{ce} gives $\Phi_1\sim\s^2M_p/m^2$. 

It is interesting to compare \eq{er} with the equation (8.32) of Kolb and Turner \cite{kt}, which gives the evolution of the radiation energy density extracted from the inflaton  by a constant decay rate $\Gamma$. It is easy to see that both equations agree for $\Gamma\sim \s^2/m$, which up to a numerical factor is the total decay rate corresponding to the interaction \eq{int2}. 

Using the value of $\Phi_1$ determined above, the Hubble constant $H_1$ at the end of the decay can be found from \eq{sol} as $H_1\sim \s^2/m$. Since our aim is to calculate the deviation in the number density of particles produced in a Hubble volume at the end of reheating we choose
\be
\e=a_1 H_1\sim a_1\fr{\s^2}{m}.
\ee
For $m\gg\s$ and $k$ in the decay range \eq{dr}, $\e\ll k$ and thus \eq{m1} and \eq{m2} is valid. Using \eq{bf}  and noting that at the end of the decay $\o_k\simeq k/a_1$  the deviations can be estimated  as 
\be\label{ff} 
\fr{\D \rho_\e}{<\rho_\e>}\simeq \fr{\D n_\e}{<n_\e>}\sim\fr{\s}{M_p}. 
\ee 
Assuming $m=10^{-6}M_p$ and $\s=10^{-7}m$ (to satisfy the requirement $m^2/M_p>\s$ imposed by the stationary phase approximation), one finds that the relative deviation is of the order of $10^{-13}$, which is very small compared to the deviation found after preheating. 

The reheating temperature in this model can be estimated from the energy density at the end of the decay (see e.g. \cite{erh1,kofrev}) 
\be
T_R^4\sim m^2\Phi_1^2\sim\left[\fr{\s^2M_p}{m}\right]^2,
\ee 
which gives a very low reheating temperature $T_R\sim 10^{9}$ GeV. Therefore  the decay actually happens rather slowly giving  a large spectrum of created particles  to be averaged over and a very small $\e$. This explains both the smallness of the deviation and the contrast with the quantum mechanical result \eq{dn}, which gives an extremely large deviation in perturbation theory. Note that the reheating temperature is independent of the initial value of the amplitude $\Phi_0$, as it should be. 

To determine the deviation for larger reheating temperatures, we first note that in the perturbative regime $|\a_k|\simeq 1$ and \eq{m1} becomes
 \be\label{g1}
\fr{\D n_\e}{<n_\e>}\simeq \,\fr{\sqrt{2}\,\e^{3/2}}{(\int d^3 k |\b_k|^2)^{1/2}}.
\ee 
The integral in \eq{g1} can  approximately be estimated in terms of the reheating temperature as $T_R^4/m$, where we set $a_1=1$. Similarly the Hubble scale at the end of reheating is $H\simeq T_R^2/M_p=\e$. Therefore, the relative deviation in the number density of particles produced in a Hubble volume is
\be
\fr{\D n_\e}{<n_\e>}\sim\fr{T_R}{M_p}\sqrt{\fr{m}{M_p}}.
\ee
For $T_R=10^{15}$ GeV and $m=10^{-6}M_p$ the relative deviation becomes $10^{-7}$. Thus the deviation appeared in perturbation theory is generically smaller than the one encountered in the parametric resonance regime. This result can be explained by the fact that preheating occurs suddenly in a very short time (i.e. explosively) and in a comparatively narrow band as measured by the Hubble scale. As a result  the number of the modes to be averaged over is small as compared to the perturbative regime, which lessens the statistics and gives a larger total deviation. 

\section{Conclusions}\label{sec5}

In this paper, we study the particle creation effects in the context of quantum fields in time dependent external backgrounds. First considering the quantum mechanical harmonic oscillator with a time dependent   frequency, which is the prototype of the field theory problem, we observe that the number of created quanta largely deviates about its mean value. Armed with this observation, we use a complete orthonormal family of localized wave packets to calculate the deviations in the field theory side.

The wave packets that  we introduce following \cite{haw} are defined with a fixed momentum scale $\e$. Roughly speaking, they carry discrete momentum values which are split up by $\e$ and are localized in a region of size $1/\e$. For small enough $\e$, they also become eigenvectors of the Hamiltonian. Therefore, these wave packets form a legitimate basis in the Hilbert space as good as the Fourier modes. Indeed, as discussed in section \ref{sec3}, it is a lot easier to interpret them physically as long as the measurements are concerned. 

To understand $\e$ dependence of the relative deviations assume that particles are produced around momentum $k$ in a band having a width $\D k$ with equal average production. For each wave packet in the band, the deviation in the number of created quanta is given by the quantum mechanical result \eq{dn}. There are however nearly $k^2 \D k/\e^3$ number of different wave packets in this band. Therefore, the relative deviation of the total number of created particles should decrease like $\e^{3/2}/(k^2\D k)^{1/2}$. The deviation found in preheating \eq{fd} can easily be re-derived by noting that $\D k\sim k \sim \sqrt{gm\Phi_0}$ and $\e\sim m$. Since the decay is very quick the dependence on the expansion of the universe is very weak in this case. However, in perturbation theory the decay is very slow and this naive estimate should be refined. Note also that decreasing $\e$ is equivalent to observing a larger space and thus it is natural to get a small deviation. In an experimental setup, the $\e$ parameter should be set  by details of the measurement. On the other hand, in an expanding universe the Hubble distance gives a natural scale. 

Applying this general framework to inflation we find that there emerges density inhomogeneities on Hubble length scales at the end of reheating. Depending on the details of the decay, the relative order of these inhomogeneities can be as large as $10^{-3}$ in preheating and $10^{-7}$ in perturbation theory. Note that local interactions are incapable of restoring homogeneity since they cannot operate on distances larger than the Hubble radius. Moreover, these estimates actually give a lower bound since the deviation in the number of  each particular mode can be larger. On the other hand, it is much safer to estimate the deviation in the total number of created particles since for small $\e$ the nearby momentum levels become physically indistinguishable.

It would be interesting to see whether these density inhomogeneities affect cosmological events following reheating like baryogenesis. It is also important to determine if they switch to the non-linear regime and grow in time due to gravitational collapse (note that since they are different than the usual Fourier modes, the standart arguments cannot be applied here). If this does not happen, interactions can gradually restore homogeneity as the Hubble radius grows, but the time required for full equilibrium might considerably be large which would reduce the reheating temperature. In any case, due to the quantum mechanical nature of the particle creation process in reheating, we find that the hot stage does not start with the perfect homogeneity attained after inflation.

\acknowledgments{This work is partially supported by Turkish Academy of Sciences via Young Investigator Award Program (T\"{U}BA-GEB\.{I}P).}

\end{document}